\documentclass[a4paper]{article}
\usepackage{tikz}
\usetikzlibrary{shapes.geometric, arrows}
\tikzstyle{layers} = [rectangle, rounded corners, minimum width=3cm, minimum height=0.5cm,text centered, draw=black]
\tikzstyle{arrow} = [thick,->,>=stealth]
\usepackage{multirow}
\usepackage{INTERSPEECH2022}
\usepackage{url}

\title{
Overlapped speech and gender detection with WavLM pre-trained features}
\name{Martin Lebourdais$^1$, Marie Tahon$^1$, Antoine Laurent$^1$, Sylvain Meignier$^1$ }
\address{
  $^1$LIUM, Le Mans Université, France
}
\email{[firstname].[name]@univ-lemans.fr}

\begin{document}

\maketitle
\begin{abstract}
This article focuses on overlapped speech and gender detection in order to study interactions between women and men in French audiovisual media (Gender Equality Monitoring project).
In this application context, we need to automatically segment the speech signal according to speakers gender, and to identify when at least two speakers speak at the same time.
We propose to use WavLM model which has the advantage of being pre-trained on a huge amount of speech data, to build an overlapped speech detection (OSD) and a gender detection (GD) systems.
In this study, we use two different corpora. The DIHARD III corpus which is well adapted for the OSD task but lack gender information. The ALLIES corpus fits with the project application context.
Our best OSD system is a Temporal Convolutional Network (TCN) with WavLM pre-trained features as input, which reaches a new state-of-the-art F1-score performance on DIHARD.
A neural GD is trained with WavLM inputs on a gender balanced subset of the French broadcast news ALLIES data, and obtains an accuracy of $94.9\%$.
This work opens new perspectives for human science researchers regarding the differences of representation between women and men in French media.


%
\end{abstract}

\section{Introduction}
Gender equality in audiovisual media is a societal concern of major importance.
Gender representation is mainly studied with statistical analysis carried on various features such as speaker role, gender, speaking time, extracted directly from manual annotations~\cite{csa_representation_2020}.
However, due to the high cost of manual annotations, such studies have a limited impact in terms of covered topics and amount of data.
Recently, systems have been developed in order to automatically extract some descriptive features,
like 
gender~\cite{doukhan_open-source_2018}.
However, gender information alone is not enough to study interactions between women and men in audiovisual media and additional information such as overlapping speech is needed.
In the framework of the French GEM (Gender Equality Monitoring) ANR project, 
we aim at automatically detecting interruptions in regards to the gender of the concurring speakers.
Interruption being a subjective notion, we use the presence of overlapped speech as a proxy to reduce the number of candidate areas. 
In this article, we thus focus on both overlapped speech detection (OSD), and gender detection (GD) from speech.

As aforementioned, OSD is interesting in the scope of our project, but has also a wide range of applications in automatic speech processing.
Indeed, it consists in extracting speech segments with at least two concurrent speakers. 
Therefore, it is widely used as a pre-treatment to ensure processing of single-speaker speech segments. As shown in~\cite{garcia_perera_speaker_2020} the presence of untreated overlapped speech degrade the performances in diarisation (speaker segmentation and clustering which answers the question ``who speaks when''). 
It thus proved to be a fair improvement to handle them~\cite{bullock_overlap-aware_2020}. It is also a useful task in automatic speech recognition, as most systems assume that they are fed with single-speaker utterance. Overlapped speech thus are an important source of error~\cite{cetin_analysis_2006}.

GD is usually considered as a 2-classes classification task (male or female) of speech segments.
This binary approach is motivated with physiological reasons, for instance the fact that men usually have lower pitch than women.
However, the fact that a system can only return two values can raise some social issues, for example in the case of non-binary gender definition.
In the data used in our work, gender annotations follows this binary approach, thus our proposed systems will be able to return only two values.

The paper is organised in the following way : Section~\ref{sec:rel-work} presents current and past works on overlapped speech detection, gender detection, and the pre-trained feature extractor fundamental in our experiment. Section~\ref{sec:corpus} describes the corpora and the features used in our experiments. Sections~\ref{sec:osd} and~\ref{sec:gd} present our overlapped speech detection and gender detector systems with pre-trained features in input. Finally, the last section discusses the results and shows possible applications in the context of audiovisual media.

\section{Related works}\label{sec:rel-work}
\subsection{Overlapped Speech detection}
OSD is a sequence-to-sequence classification task, \textit{i.e.} the automatic system has to return a numeric sequence given a speech segment in input.
Usually the output frame rate is set to 10~ms, and the output is binary
with 1 meaning the presence of overlapped speech and 0 otherwise. 
Some works combined an overlapped speech detector with a voice activity detector or a speaker counter, thus using more than 2 classes~\cite{jung_three-class_2021,bredin:hal-03257524}.

Multiple representations of the speech signal are used for automatic audio signal processing. Low-level acoustic features are used for a long time, such as Mel Frequency Cepstral Coefficients (MFCCs) and Mel Filter Banks.
These features both use a Mel scale, which is close to human perception of sound, and still bring fairly good results with low calculation times.

Improvements have been made on acoustics features with for example the pyknogram, an enhanced time-frequency representation~\cite{shokouhi_teagerkaiser_2017}. 

With the extensive use of neural networks, the current trend is to get rid of hand-crafted acoustic features and let the network learn its own best acoustic representation given the task.
For example, SincNet~\cite{ravanelli_speaker_2018} network originally developed for speaker identification, is able to learn a speaker representation directly from the time domain.
Such a representation has been proven to be very efficient for OSD~\cite{bullock_overlap-aware_2020}.
$x$-vectors~\cite{snyder_x-vectors_2018}, \textit{i.e.} considered as speaker embeddings, are trained for speaker identification, but are also efficient for OSD~\cite{malek_voice-activity_2020}.

Most of the current OSD systems use sequence-to-sequence neural architectures, mainly recurrent networks including LSTMs (Long Short-Term Memory)~\cite{geiger_detecting_2013}. This architecture is a relevant choice due to its capacity to deal with variable length sequences. The bi-directional variant (BiLSTM), which relies on the past and future context,
brings an improvement as overlapped speech detection depends on the context.

Convolutional neural networks (CNN) have been recently used in audio and speech processing. Following this trend, OSD systems based on convolutional layers are becoming frequent~\cite{kunesova_detection_2019}, granting results as good as the one obtained with recurrent layers, with smaller training duration. 
Some OSD systems combine recurrent and convolutional layers to improve performances~\cite{jung_three-class_2021}.
Finally, the Temporal Convoluted Network (TCN) originally developed for sequence modelling~\cite{Bai2018AnEE} have been adapted for speaker counting in overlapped speech~\cite{cornell_detecting_2020}.
Indeed, this network combines a good representation of the low-level context with the first convolutional layer, and exploits a long context provided by stacked dilated convolutional layers. 
Speaker counting being reasonably close but harder compared to overlapped speech detection, we hope that the issue presented in~\cite{cornell_detecting_2020} on high number of concurring speakers will not apply to our task. 

\subsection{Gender recognition}
Gender detection (GD) from speech is usually considered as a 2-classes classification task where the model returns either `female' or `male'.
Model performances are evaluated in terms of accuracy.
This task is usually considered as well-defined and almost resolved task.
Indeed, different models have been used from the 90s (two HMM~\cite{parris_language_1996} in 1996, or SVM~\cite{bocklet_age_2008} in 2008), and all of them have reached accuracy above 90\%. 
More recently, neural networks have been used to predict gender from voice~\cite{doukhan_open-source_2018}. 
The lack of common corpus, combined with good results in each evaluation, makes it difficult to assess the degree of improvement of these types of systems.

\subsection{Pre-trained features for audio segmentation}

Recently, following the success of the language model BERT~\cite{Devlin2019BERTPO} on textual data representation, various models which learn audio data representation have been developed. 
The objective is to train a system to a single task requiring an extensive representation of the underlying audio data. 
Once the network is trained, the last few layers are then removed to get a system building an extensive vectorial representation of audio data, which can then be used as input features. 
Overlapped speech can be considered as rare in speech signals (less than 10\% in duration), therefore we lack real data to train OSD models. 
Such pre-trained features can help to mitigate the small amount of overlapped speech data, because they were learnt using huge quantity of data.
Our work is based on the new feature extractor created by Microsoft, named WavLM~\cite{Chen2021WavLM}. 
This system is a new self-supervised system built with transformer blocks trained on Mix94k, a corpus of 94k hours drawn from LibriLight, VoxPopuli and GigaSpeech. WavLM learns to represent speech by masking a part of the signal and trying to predict the hidden part. 
On this aspect, this system is similar to the self-supervised systems HuBERT~\cite{hsu_hubert_2021} and wav2vec2.0~\cite{baevski_wav2vec_2020}.
During the training, artificial overlap has been implemented to augment the number of data by summing two audio segments. This makes WavLM particularly suitable for overlapped speech detection as it already seen artificially overlapped speech during the training phase. This is the major reason, we chose this feature extractor for our speech processing tasks.
In order to be able to represent the diversity of the speech data used to train WavLM, the network architecture involves a lot of parameters.
For example, its large variant, which reaches the best performances on the SUPERB benchmark~\cite{yang_superb_2021} has 316.62~M parameters.
Consequently, such a model is very costly to train, in time, energy and money. Fortunately, pre-trained models are available on torch hub~\footnote{https://pytorch.org/hub/} in all of its variants. It is therefore usable by the vast majority of the community.


\section{Corpus and features}\label{sec:corpus}
\subsection{Corpus}

This section presents the two corpus used in our study.
DIHARD corpus is used to train and evaluate OSD systems, while ALLIES is used to train and evaluate GD systems.
Table~\ref{tab:corpus} summarizes the characteristics of these databases.

\begin{table}[ht]
    \centering
    \caption{Characteristics of the annotated corpora: total duration, overlap and gender proportion in duration, number of speakers (female speakers).}
    \begin{tabular}{|l|c|c|c|c|}
    \hline
        \textbf{Corpus} & \textbf{Dur.} & \textbf{Overlap} & \textbf{Female} & \textbf{\#Spks (F)}\\\hline
        DIHARD & 34h & 11.6\%  & NA & NA\\\hline
        ALLIES & 307h & 3.2\% & 27.3\% & 5711 (2001)\\\hline
        ALLIES-G & 18h  & NA  & 50.0\% & 1576 (798)\\\hline
        
    \end{tabular}
    
    \label{tab:corpus}
\end{table}

\textbf{The DIHARD corpus}~\cite{ryant2021dihard} is the corpus provided for the eponymous challenge in 2020. This corpus has been designed to contain `difficult' data, \textit{i.e.} data with various recording qualities, situations and spontaneous speech. Spontaneous speech by nature contains a lot of overlapped speech. This corpus is thus adapted to the OSD task. This corpus has been divided according to the distribution used for the evaluation campaign, in a train set and a development set, while the test set is provided by the organisers. 
This corpus can not be used for gender detection as it has not been annotated in gender.

\textbf{The ALLIES corpus} is a French broadcast media corpus that extends previous speech evaluation campaigns~\cite{larcher:hal-03262914}.
This corpus is composed of different types of shows (broadcast news, interviews, debates) which have been partially segmented in terms of speaker and gender, thus making it suitable to our gender detection task. 
The ALLIES annotated partition represents \~307h of gender unbalanced speech: female speech represents only 27.3\% of the total duration.
To ensure a minimal data bias for gender detection, we extract a balanced subset of this corpus, referred as ALLIES-G in Table~\ref{tab:corpus}.
The limiting number of speakers is the number of female speakers (798). The eventual cross-show presence of speakers with different names has not been treated.
From ALLIES reference segmentation, we extract single speaker speech segments.
We then split the speakers in train and test sets with the following rules:
\begin{itemize}
    \item The intersection between train, dev and test sets must be null.
    \item There must be 40 females and 40 males in test set.
    \item There must be the same number of males and females in the train set. 
\end{itemize}
From these selected speakers, we select 4000 1s segments for test set and 60000 1s-segments for train set by balancing the number of speech segments per speaker~\footnote{Available on \url{https://git-lium.univ-lemans.fr/mlebourdais/corpora/}}.

To be able to further evaluate our systems on full shows, 11 audio files have been discarded from ALLIES-G.
These files have been manually chosen because they contain a lot of interactions and different genders. This test set consists of three 1h debate shows from 2011 to 2014 produced by the French TV media LCP.

\subsection{Features}

In our experiments, two types of features are used: low-level features MFCCs and WavLM pre-trained features.

\textbf{MFCCs} are extracted on 2s-audio segments to serve as a baseline. 20 MFCCs completed by the deltas and the deltas second without the energy for a total vector of dimension 59, are extracted every 10ms on a 30ms window. 

\textbf{WavLM: }
The second set of features is extracted with WavLM~\cite{Chen2021WavLM}.
For OSD, we used the large version of WavLM~\footnote{model is available on torch.hub at wavlm\_large in s3prl/s3prl} which returns 1024-dimension vectors per frame, without fine-tuning the model. 
A segment of 2 seconds at 16~kHz is given in input of our model.  
A linear layer added on top of WavLM enables to return a 200 samples sequence, aligned with our reference ($1$ in presence of overlapped speech, $0$ in other case). 
GD is usually considered as an `easy' task. Therefore, we decided to use a smaller model, also trained on the same Mix94k, called `Base-plus' in  WavLM nomenclature. It returns 768-dimension vectors. 

\section{Overlapped speech detector (OSD)}\label{sec:osd}

\subsection{Architectures}
\begin{figure}[ht]
    \centering
    
    \begin{tikzpicture}[node distance=0.8cm]
        \node (rnn1) [layers] {BiLSTM 128};
        \node (rnn2) [layers,above of=rnn1] {BiLSTM 128};
        \node (lin1) [layers,above of=rnn2] {Linear 128};
        \node (lin2) [layers,above of=lin1] {Linear 128};
        \node (lin3) [layers,above of=lin2] {Linear 2};
        
        \draw [arrow] (rnn1) -- (rnn2);
        \draw [arrow] (rnn2) -- (lin1);
        \draw [arrow] (lin1) -- (lin2);
        \draw [arrow] (lin2) -- (lin3);

    \end{tikzpicture}
    \caption{Recurrent Overlap Speech Detector (ROSD) network.}
    \label{fig:rnn}
\end{figure}
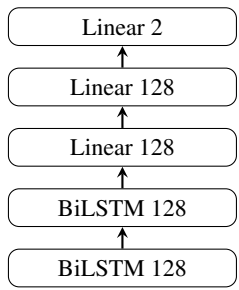

First, a Recurrent Overlap Speech Detector (ROSD, Fig.~\ref{fig:rnn}), adapted from pyannote OSD system~\cite{bredin_pyannoteaudio_2020}, contains two 128-dimensional BiLSTM layers followed by two 128-dimensional linear layers and a 2-dimensional output layer representing overlapped speech and non overlapped speech.

The second system (referred as TCN) is the TCN architecture developed for speaker counting in~\cite{cornell_detecting_2020}. 
In our case, the output layer is a binary classifier which returns overlapped speech and non overlapped speech classes.
Both architectures are trained with a cross-entropy loss during a maximum of 120 epochs.

\subsection{OSD system results}\label{sec:results}

\begin{table}[ht]
\centering
\caption{OSD results on DIHARD3 test. F1-score, Precision and Recall in \%}
\begin{tabular}{|ccc|c|c|c|}
\hline
\multicolumn{1}{|c|}{\textbf{Input}} & \multicolumn{1}{c|}{\textbf{Archi}} & \textbf{Param} & \textbf{Prec} & \textbf{Recall} & \textbf{F1-score} \\ \hline
\multicolumn{1}{|c|}{MFCC}           & \multicolumn{1}{c|}{ROSD}          & 0.638 M        & 34.2         & 60.8           & 43.8             \\ \hline
\multicolumn{1}{|c|}{MFCC}           & \multicolumn{1}{c|}{TCN}            & 0.268 M        & 46.6         & 59.8           & 52.4             \\ \hline
\multicolumn{1}{|c|}{WavLM}          & \multicolumn{1}{c|}{ROSD}          & 1.647 M        & 61.0         & 63.6           & 62.3             \\ \hline
\multicolumn{1}{|c|}{WavLM}          & \multicolumn{1}{c|}{TCN}            & 0.352 M        & 60.1         & 67.1           & \textbf{63.4}             \\ \hline
\multicolumn{3}{|c|}{Bredin \textit{et al.}~\cite{bredin:hal-03257524}}                                                               & 57.2         & 62.8           & 59.9             \\ \hline
\end{tabular}
\label{tab:ov_test_dihard}
\end{table}

The experiment is conducted on DIHARD3 for ROSD and TCN architectures combined with MFCC and WavLM inputs.
Table\ref{tab:ov_test_dihard} shows F1-score, precision and recall on the test set.
The state of the art performance for OSD is claimed by~\cite{bredin:hal-03257524} with a F1-score of 59.9\%.
TCN and ROSD systems based upon WavLM significantly beats this result with respectively 63.4\% and 62.3\% of F1-score. 

More precisely, WavLM features (63.4\% with TCN) seems to better represent overlapped speech than MFCC (52.4\% with TCN) with an absolute gain of almost 11 points. A small advantage (1.1 in absolute) for TCN is observed over ROSD with WavLM features.

The second advantage of using TCN with high-dimensional inputs such as WavLM is that the first convolutional layer of this network reduces rapidly the number of dimensions, thus limiting drastically the overall parameters.

\section{Gender classification}\label{sec:gd}
\subsection{Architectures}
The gender classification backbone is a recurrent network with one 64-dimensional LSTM layer.
Two different approaches are tested which only differ by the last output layer.

The first approach referred as GD1 outputs two classes (female, male). A 2-dimensional output linear layer is added to the backbone and summed for each class before applying a softmax. The argmax of the softmax output gives the final predicted gender.
In this approach the model is trained with a cross-entropy loss.
The main drawback of this approach is that, in case of overlapped speech, it can predict only one gender.

To cope with this issue, we propose a second approach based on two independent models: the first predicts the presence of male, while the second predicts the presence of female.
For each, a linear layer outputs a numerical value, summed at the segment level.
The final prediction, is the argmax between the outputs of the two values given by the two models.
This approach is considered as a regression task, therefore a RMSE loss is used to train the two models.
This second approach is called GD2 in the following tables. 

In preliminary experiments, we have shown that WavLM outperforms MFCCs input features in the same manner than for OSD. Therefore, only WavLM results are reported.
All the systems are trained on the balanced ALLIES-G train corpus. 
Except the number of epochs, no hyper parameters needs to be tuned on a development set.
The final models are obtained with 2 epochs only as the loss function do not significantly improves after.

\subsection{GD system results}

\begin{table}[ht]
\centering
\caption{Accuracy results (\%) of GD systems on ALLIES-G test}
\begin{tabular}{|c|c|c|c|}
\hline
                                         GD1                 & 94.9        & 97.8          & 92.1          \\ \hline 
                                          GD2        &    94.4     &    98.0       &   90.8        \\ \hline
\end{tabular}
\label{tab:gender_test}
\end{table}



Our systems are evaluated with a global accuracy (Acc) that counts the number of well-classified samples over the total number of samples, and one accuracy per gender (Acc\_M, Acc\_F).
The results (see Table~\ref{tab:gender_test}) of our systems all reach results above 90\% of accuracy consistent with state of the art.
We can notice that even when the data bias is strongly limited, female accuracy is generally lower than male accuracy, in the same way as most of speaker recognition tasks.
GD2 system gives slightly lower performances than GD1, however GD2 has the advantage of being tunable with a threshold in order to better balance the accuracy between classes.
\subsection{Analysis of GD errors according to pitch}
\begin{figure}[ht]
    \centering
    \includegraphics[width=\linewidth]{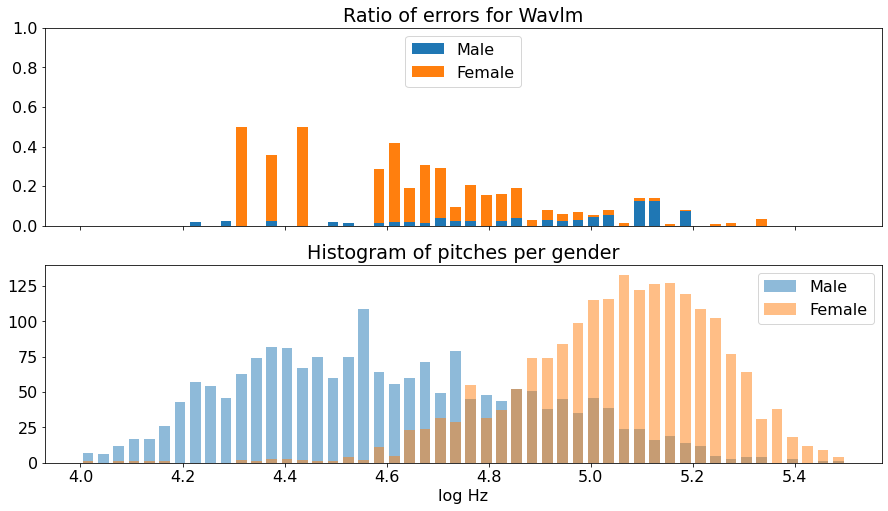}
    \caption{Gender detection errors normalized by the total number of samples with respect to voice pitch in $\log F0$}
    \label{fig:gender_rep}
\end{figure}

We noticed that the GD accuracy is extremely variable according to the speakers.
We hypothesize that both systems build their decision on implicit pitch information.

The Figure~\ref{fig:gender_rep} shows the distribution of errors with respect to the pitch extracted on all segments using YIN algorithm~\cite{de_cheveigne_yin_2002}.

We expected that most of the errors occur where the two distributions overlap, however we observe that the errors mainly occurs in the distribution tails.
For example, the 5\% female speech segments with the lowest pitch ($\log F0$ between 4 and 4.69), those the most confused with the male distribution, only have Acc\_F=76.9\%.
In comparison, the female segments in the middle of the distribution reach Acc\_F= 96.8\%, and 97.8\% for those in the 5\% highest part of the distribution.
The same tendency is also observed for male speakers: 96.4\% for the highest 5\%, 98.3\% for the middle of the distribution, 100\% for the not mixed part. 
From this analysis, we confirm that GD model decisions are affected by pitch related information but also by additional cues to be further explored.

\section{Discussions and perspectives}\label{sec:exp_genderfull}

In the medium term, the objective is to automatically detect genders into overlapped areas. 
We propose to evaluate a combined overlapped speech detector and gender detector system on one of the complete debate show from ALLIES that have never been seen in training phases. 

In this preliminary study, our OSD system has not been adapted to ALLIES data, therefore we will only consider reference overlap segments and gender scores predicted by the two gender models (GD2 approach).
GD2 output values are computed over a sliding window of 1 second with a 10ms step. 

\begin{figure}
    \centering
    \includegraphics[width=\linewidth]{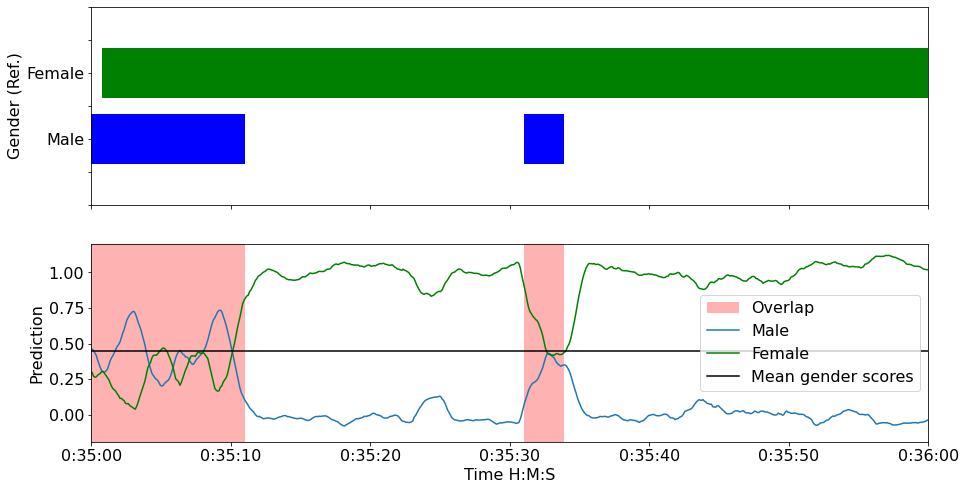}
    \caption{Gender segmentation reference (top), gender prediction scores (bottom) and reference overlaps (red rectangles) on a full show (one minute excerpt)}
    \label{fig:full_pred}
\end{figure}
Figure~\ref{fig:full_pred} shows an example for 1 minute of speech drawn from a debate (ALLIES corpus).
The predicted gender scores are plotted in the bottom panel as well as the reference overlaps (red rectangles). The top panel shows the reference gender segmentation.
From 35'10'' to the end, a female speaker has been identified (the green curve is over the blue one). 
From the beginning to 35'10'' and around 35'30'' two overlap segments are present in which both male and female are speaking. The male and female scores are close to an absolute male and female average. We hypothesize that it indicates a probable presence of both genders. 

The ALLIES corpus is dedicated to single and cross speaker diarization and speaker identification. Most of the speakers are precisely identified by forenames and family names, while the overlapped speech boundaries and gender annotations are less accurate. This preliminary study shows that references need to be manually verified before proposing a joint evaluation framework for overlapped speech and gender detection.  

 

\section{Conclusions}
In the framework of GEM project, we study overlapped speech detection and gender detection tasks. We present two systems based on WavLM and neural network architectures (TCN, LSTM and biLSTM). Experiments, conducted on the DIHARD3 and ALLIES corpora, overtake the state of the art for overlapped speech detector whereas the gender detector reaches an accuracy of 94.9\% on a gender balanced dataset. This therefore concludes that WavLM based architecture are versatile and competitive with standard features.
Moreover, pitches and dynamics of gender scores show promising results, thus motivating the further development of a joint gender and overlapped speech detector.


\section{Acknowledgements}

Our research is funded by the French national agency project GEM, Gender Equality Monitoring (ANR-19-CE38-0012). Experiments were conducted using SideKit toolkit~\cite{larcher:hal-01433157} dedicated to speaker recognition.

\bibliographystyle{IEEEtran}

\bibliography{mybib}

\begin{thebibliography}{10}
\providecommand{\url}[1]{#1}
\csname url@samestyle\endcsname
\providecommand{\newblock}{\relax}
\providecommand{\bibinfo}[2]{#2}
\providecommand{\BIBentrySTDinterwordspacing}{\spaceskip=0pt\relax}
\providecommand{\BIBentryALTinterwordstretchfactor}{4}
\providecommand{\BIBentryALTinterwordspacing}{\spaceskip=\fontdimen2\font plus
\BIBentryALTinterwordstretchfactor\fontdimen3\font minus
  \fontdimen4\font\relax}
\providecommand{\BIBforeignlanguage}[2]{{%
\expandafter\ifx\csname l@#1\endcsname\relax
\typeout{** WARNING: IEEEtran.bst: No hyphenation pattern has been}%
\typeout{** loaded for the language `#1'. Using the pattern for}%
\typeout{** the default language instead.}%
\else
\language=\csname l@#1\endcsname
\fi
#2}}
\providecommand{\BIBdecl}{\relax}
\BIBdecl

\bibitem{csa_representation_2020}
``La représentation des femmes à la télévision et à la radio,'' Conseil
  supérieur de l'audiovisuel, Tech. Rep., 2021.

\bibitem{doukhan_open-source_2018}
D.~Doukhan, J.~Carrive, F.~Vallet, A.~Larcher, and S.~Meignier, ``An
  {Open}-{Source} {Speaker} {Gender} {Detection} {Framework} for {Monitoring}
  {Gender} {Equality},'' in \emph{{ICASSP}}, Apr. 2018, pp. 5214--5218, iSSN:
  2379-190X.

\bibitem{garcia_perera_speaker_2020}
L.~P. Garcia~Perera, J.~Villalba, H.~Bredin, J.~Du, D.~Castan, A.~Cristia,
  L.~Bullock, L.~Guo, K.~Okabe, P.~S. Nidadavolu, S.~Kataria, S.~Chen,
  L.~Galmant, M.~Lavechin, L.~Sun, M.-P. Gill, B.~Ben-Yair, S.~Abdoli, X.~Wang,
  W.~Bouaziz, H.~Titeux, E.~Dupoux, K.~A. Lee, and N.~Dehak,
  ``\BIBforeignlanguage{en}{Speaker {Detection} in the {Wild}: {Lessons}
  {Learned} from {JSALT} 2019},'' in \emph{\BIBforeignlanguage{en}{The
  {Speaker} and {Language} {Recognition} {Workshop} ({Odyssey} 2020)}}, 2020,
  pp. 415--422.

\bibitem{bullock_overlap-aware_2020}
L.~Bullock, H.~Bredin, and L.~P. Garcia-Perera, ``Overlap-{Aware}
  {Diarization}: {Resegmentation} {Using} {Neural} {End}-to-{End} {Overlapped}
  {Speech} {Detection},'' in \emph{{ICASSP}}, Barcelona, Spain, 2020, pp.
  7114--7118.

\bibitem{cetin_analysis_2006}
O.~Çetin and E.~Shriberg, ``Analysis of overlaps in meetings by dialog
  factors, hot spots, speakers, and collection site: insights for automatic
  speech recognition,'' in \emph{{Interspeech}}, Pittsburgh, USA, 2006, pp.
  paper 1915--Mon2A2O.6.

\bibitem{jung_three-class_2021}
J.-w. Jung, H.-S. Heo, Y.~Kwon, J.~S. Chung, and B.-J. Lee,
  ``\BIBforeignlanguage{en}{Three-{Class} {Overlapped} {Speech} {Detection}
  {Using} a {Convolutional} {Recurrent} {Neural} {Network}},'' in
  \emph{\BIBforeignlanguage{en}{Interspeech 2021}}.\hskip 1em plus 0.5em minus
  0.4em\relax ISCA, Aug. 2021, pp. 3086--3090.

\bibitem{bredin:hal-03257524}
H.~Bredin and A.~Laurent, ``{End-to-end speaker segmentation for overlap-aware
  resegmentation},'' in \emph{{Interspeech}}, Brno, Czech Republic, 2021.

\bibitem{shokouhi_teagerkaiser_2017}
N.~Shokouhi and J.~H.~L. Hansen, ``\BIBforeignlanguage{en}{Teager–{Kaiser}
  {Energy} {Operators} for {Overlapped} {Speech} {Detection}},''
  \emph{\BIBforeignlanguage{en}{IEEE/ACM Transactions on Audio, Speech, and
  Language Processing}}, vol.~25, no.~5, pp. 1035--1047, May 2017.

\bibitem{ravanelli_speaker_2018}
M.~Ravanelli and Y.~Bengio, ``Speaker {Recognition} from {Raw} {Waveform} with
  {SincNet},'' \emph{Speech and Language Technology {SLT}}, pp. 1021--1028,
  2018.

\bibitem{snyder_x-vectors_2018}
D.~Snyder, D.~Garcia-Romero, G.~Sell, D.~Povey, and S.~Khudanpur,
  ``\BIBforeignlanguage{en}{X-{Vectors}: {Robust} {DNN} {Embeddings} for
  {Speaker} {Recognition}},'' in
  \emph{\BIBforeignlanguage{en}{{ICASSP}}}.\hskip 1em plus 0.5em minus
  0.4em\relax Calgary, AB: IEEE, 2018, pp. 5329--5333.

\bibitem{malek_voice-activity_2020}
J.~Málek and J.~Žďánský, ``Voice-{Activity} and {Overlapped} {Speech}
  {Detection} {Using} x-{Vectors},'' in \emph{Text, {Speech}, and
  {Dialogue}}.\hskip 1em plus 0.5em minus 0.4em\relax Cham: Springer
  International Publishing, 2020, pp. 366--376.

\bibitem{geiger_detecting_2013}
J.~Geiger, F.~Eyben, B.~Schuller, and G.~Rigoll, ``Detecting overlapping speech
  with long short-term memory recurrent neural networks,'' in
  \emph{Interspeech}, Lyon, France, 2013, pp. 1668--1672.

\bibitem{kunesova_detection_2019}
M.~Kunešová, M.~Hrúz, Z.~Zajíc, and V.~Radová,
  ``\BIBforeignlanguage{en}{Detection of {Overlapping} {Speech} for the
  {Purposes} of {Speaker} {Diarization}},'' in
  \emph{\BIBforeignlanguage{en}{Speech and {Computer}}}, ser. Lecture {Notes}
  in {Computer} {Science}.\hskip 1em plus 0.5em minus 0.4em\relax Cham:
  Springer International Publishing, 2019, pp. 247--257.

\bibitem{Bai2018AnEE}
S.~Bai, J.~Z. Kolter, and V.~Koltun, ``An empirical evaluation of generic
  convolutional and recurrent networks for sequence modeling,'' \emph{arXiv},
  vol. abs/1803.01271, 2018.

\bibitem{cornell_detecting_2020}
S.~Cornell, M.~Omologo, S.~Squartini, and E.~Vincent, ``Detecting and
  {Counting} {Overlapping} {Speakers} in {Distant} {Speech} {Scenarios},'' in
  \emph{{Interspeech}}, Shanghai, China, 2020, pp. 3107--3111.

\bibitem{parris_language_1996}
E.~Parris and M.~Carey, ``Language independent gender identification,'' in
  \emph{{ICASSP}}, 1996, pp. 685--688.

\bibitem{bocklet_age_2008}
T.~Bocklet, A.~Maier, J.~G. Bauer, F.~Burkhardt, and E.~Nöth, ``Age and gender
  recognition for telephone applications based on {GMM} supervectors and
  support vector machines,'' \emph{ICASSP}, pp. 1605--1608, 2008.

\bibitem{Devlin2019BERTPO}
J.~Devlin, M.-W. Chang, K.~Lee, and K.~Toutanova, ``Bert: Pre-training of deep
  bidirectional transformers for language understanding,'' in \emph{NAACL},
  2019.

\bibitem{Chen2021WavLM}
S.~Chen, C.~Wang, Z.~Chen, Y.~Wu, S.~Liu, Z.~Chen, J.~Li, N.~Kanda,
  T.~Yoshioka, X.~Xiao, J.~Wu, L.~Zhou, S.~Ren, Y.~Qian, Y.~Qian, J.~Wu,
  M.~Zeng, and F.~Wei, ``Wavlm: Large-scale self-supervised pre-training for
  full stack speech processing,'' \emph{arXiv}, 2021.

\bibitem{hsu_hubert_2021}
W.-N. Hsu, B.~Bolte, Y.-H.~H. Tsai, K.~Lakhotia, R.~Salakhutdinov, and
  A.~Mohamed, ``\BIBforeignlanguage{en}{{HuBERT}: {Self}-{Supervised} {Speech}
  {Representation} {Learning} by {Masked} {Prediction} of {Hidden} {Units}},''
  \emph{\BIBforeignlanguage{en}{arXiv}}, Jun. 2021.

\bibitem{baevski_wav2vec_2020}
A.~Baevski, H.~Zhou, A.~Mohamed, and M.~Auli, ``\BIBforeignlanguage{en}{wav2vec
  2.0: {A} {Framework} for {Self}-{Supervised} {Learning} of {Speech}
  {Representations}},'' \emph{\BIBforeignlanguage{en}{{arXiv}}}, Oct. 2020.

\bibitem{yang_superb_2021}
S.-W. Yang, P.-H. Chi, Y.-S. Chuang, C.-I.~J. Lai, K.~Lakhotia, Y.~Y. Lin,
  A.~T. Liu, J.~Shi, X.~Chang, G.-T. Lin, T.-H. Huang, W.-C. Tseng, K.-t. Lee,
  D.-R. Liu, Z.~Huang, S.~Dong, S.-W. Li, S.~Watanabe, A.~Mohamed, and H.-y.
  Lee, ``\BIBforeignlanguage{en}{{SUPERB}: {Speech} {Processing} {Universal}
  {PERformance} {Benchmark}},'' in \emph{\BIBforeignlanguage{en}{Interspeech
  2021}}, 2021, pp. 1194--1198.

\bibitem{ryant2021dihard}
N.~Ryant, P.~Singh, V.~Krishnamohan, R.~Varma, K.~Church, C.~Cieri, J.~Du,
  S.~Ganapathy, and M.~Liberman, ``The third dihard diarization challenge,'' in
  \emph{{Interspeech}}, Brno, Czechia, 2021, pp. 3570--3574.

\bibitem{larcher:hal-03262914}
A.~Larcher, A.~Mehrish, M.~Tahon, S.~Meignier, J.~Carrive, D.~Doukhan,
  O.~Galibert, and N.~Evans, ``{Speaker Embedding For Diarization Of Broadcast
  Data In The {ALLIES} Challenge},'' in \emph{{ICASSP}}, Toronto, Canada, 2021,
  pp. 5799--5803.

\bibitem{bredin_pyannoteaudio_2020}
H.~Bredin, R.~Yin, J.~M. Coria, G.~Gelly, P.~Korshunov, M.~Lavechin, D.~Fustes,
  H.~Titeux, W.~Bouaziz, and M.-P. Gill, ``Pyannote.{Audio}: {Neural}
  {Building} {Blocks} for {Speaker} {Diarization},'' in \emph{{ICASSP}},
  Barcelona, Spain, 2020, pp. 7124--7128.

\bibitem{de_cheveigne_yin_2002}
A.~de~Cheveigne and H.~Kawahara, ``\BIBforeignlanguage{en}{{YIN}, a fundamental
  frequency estimator for speech and musica)},''
  \emph{\BIBforeignlanguage{en}{J. Acoust. Soc. Am.}}, vol. 111, no.~4, p.~14,
  2002.

\bibitem{larcher:hal-01433157}
A.~Larcher, K.~A. Lee, and S.~Meignier, ``{An Extensible Speaker Identification
  SIDEKIT in Python},'' in \emph{{ICASSP)}}, Shanghai, China, 2016, pp.
  5095--5099.

\end{thebibliography}


\end{document}